*Article*

# Multilayer Thin-Film Flake Dispersion Gel for Surface-Enhanced Raman Spectroscopy

Samir Kumar*, Misa Kanagawa, Kyoko Namura, Takao Fukuoka, and Motofumi Suzuki

*Department of Micro Engineering, Graduate School of Engineering, Kyoto University, Katsura, Nishikyo, Kyoto 615-8540 Japan.*

* Correspondence: drsamirkumar2017@gmail.com

**ABSTRACT:** Ag nanorod arrays/dielectrics/mirror-structured multilayer thin-film are well known, highly sensitive surface-enhanced Raman scattering (SERS) substrates that enhance the Raman scattering cross-section by the interference of light. However, extracting biomarkers directly from human skin using these solid substrates is difficult. To overcome this problem, we propose a multilayer thin-film flake dispersion gel by centrifugal mixing of the multilayer thin-film and hydroxyethyl cellulose (HEC) gel. The multilayer thin-film was prepared by serial bideposition using the dynamic oblique angle deposition technique. The mixing process was optimized to obtain flakes of ~10 μm so that the optical properties of the multilayer film can be preserved, and there is no risk of adverse effects on humans. The SERS features of the flakes dispersion gel were tested using 4, 4'-bipyridine (BPY). The BPY molecules diffused through the highly porous gel within a few seconds, generating significant SERS signals. The multilayer film flakes dispersion gel showed a SERS signal about 20 times better than the gel-dispersed Ag nanorod arrays without a multilayer film structure. These SERS active flakes dispersion gel can be used directly on the skin surface to collect body fluids from sweat, for biomarker sensing.

## 1. INTRODUCTION

Localized plasmon resonance can be excited when the light of a specific wavelength is incident on a noble metal nanoparticle.[1,2] This enhances the electric field around the nanoparticles and enables surface-enhanced Raman scattering (SERS).[3–5] SERS has been a growing area of interest in materials science, biophysics, medical diagnostics, and molecular biology as it enables the label-free detection and identification of molecules.[6–9] With the rapid development of medical and chemical fields, efforts have recently been dedicated to the practical use of biosensors using SERS.[10–13]

A variety of SERS substrates have been reported, ranging from rough metal surfaces to fractals, nanowires, nanoparticle solutions, and periodic nanopatterns.[14–18] Ag nanorod arrays (NRA) fabricated by the dynamic oblique vapor deposition (OAD) technique are already known for their reproducibility and sensitivity with absorption peaks due to local plasmon resonance in the near-infrared region.[19,20] However, NRAs on glass substrates are not effectively utilized because they transmit or scatter about 70% of the light that excites the plasmons.[21] Consequently, Tokunaga et al. proposed an AgNRA/SiO$_2$ shape control layer (SCL)/SiO$_2$ phase control layer (PCL)/Ag mirror multilayer thin-film to efficiently utilize the light scattered in the AgNRAs.[22] A multilayered spacer layer comprising two layers performs the following functions: (1) controlling the form of nanorods, SCL, and (2) controlling the length of the optical path, PCL. By optimizing the film thickness of the PCL, the energy of the incident light was concentrated by light interference on the nanorod around the plasma resonance energy, and the reflectance became extremely small. This multilayer thin-film structure also improved the SERS intensity by 65 times compared that of the Ag NRA prepared on the glass substrate by the light confinement effect. glass and silicon are the most widely used substrates for the fabrication of plasmonic nanostructures.[23–26] Nonetheless, the solid substrates are rigid and brittle, and thus severely limit the application of SERS, such as directly collecting the trace molecules from a surface. A few reports are available on the fabrication of gel-based SERS substrates so that to remove this limitation.[27–32] However, the primary purpose of the study was to use gel to shield nanoparticle aggregates from the environment during storage and ease of handling.

Here, we report the fabrication of a multilayer thin-film flakes dispersion gel SERS sensor. We first examined a method for producing a flake dispersion gel with the desired particle size using a simple multilayer structure. Next, we used the optimized method to prepare a flakes dispersion gel using an AgNRA/SiO$_2$/SiO$_2$/Ag mirror multilayer thin-film structure and evaluated the SERS characteristics. The Ag NRA/SiO$_2$/SiO$_2$/Ag mirror gel showed excellent SERS characteristics and did not significantly compromise the SERS performance of the analyte. We propose that this flakes dispersion gel SERS sensor can be applied directly to the surface of the skin to absorb body fluids from sweat.

## 2. MATERIALS AND METHODS

### 2.1. Fabrication of Multilayer Film for Preliminary Examination of the Thin-film Flakes Dispersion Gel.

In the preliminary study for the optimization of dispersed flake size, a multilayer film with a simple structure was prepared on a 50×50 mm$^2$ mica substrate of a thickness of 15–30 μm. The schematic of the cross-section of this multilayer film is shown in Fig. 1(b). Henceforth, this sample is referred to as ML1. ML1 structure reduced the time required to produce a multilayer film because we had to fabricate a lot of samples for the optimization process. The ML1 thin-films were deposited by D.C. magnetron sputtering (manufacturer) from 50 mm SiO$_2$ and Ag targets with a purity of 99.99%. The base pressure of the chamber was $4 \times 10^{-4}$ Pa, and pure Ar was introduced as the sputtering gas at a flow rate of 9.0 sccm. The angle between the axis of the sputter gun and the substrate normal was 28°, and the distance between the center of the target and the substrate was 10 cm. The conditions for the deposition of the ML1 thin-film are given in Table 1.

Table 1. Deposition Conditions for the ML1 Thin-films.

|  | Output (W) | Pressure (Pa) | Sputtering time (s) | Film thickness (nm) |
|---|---|---|---|---|
| Ag mirror | 50 | 0.80 | 687 | 200 |
| SiO$_2$ layer | 150 | 1.0 | 1653 | 140 |
| Ag nanostructured layer | 30 | 0.80 | 34 | 10 |

### 2.2. Fabrication of NRA/Dielectric/Mirror Multilayer Film.

The dynamic OAD technique was used to fabricate AgNRA/SiO$_2$/SiO$_2$/Ag mirror multilayer thin-films. This sample is referred to as ML2. A schematic diagram of the cross-section of ML2 thin-film is shown in Fig. 1(b). Dynamic OAD is a physical vapor deposition technique in which the vapor flux is incident at a large angle with reference to the surface normal of the substrate.[33,34] The distance between the substrate and the source was 480 mm. A smooth 200 nm thick layer of Ag was deposited on a mica substrate at a deposition angle of 0°, and the average deposition rate was approximately 0.20 nm/s. SiO$_2$ was deposited at a deposition angle of 0° on top of it to prepare a PCL of 30–140 nm thickness. The average deposition rate was 0.2-0.3 nm/s. On the PCL, the SCL of SiO$_2$ with an anisotropic surface morphology was prepared by the serial bideposition (SBD) technique.[35] The SiO$_2$ tablet (99.9%) was evaporated from the electron beam source at a distance of 480 mm from the substrate at an angle of 82°, while the azimuthal angle was rapidly changed by 180° with each deposition of 20 nm at a pressure of $7 \times 10^{-3}$ Pa. A SiO$_2$-layer with an average thickness of 600 nm was obtained after 15 cycles of SBD. On the SiO$_2$ SCL, Ag was deposited obliquely at 70°, while the azimuth angle was not changed. The pressure during the Ag deposition was $3 \times 10^{-4}$ Pa, and the deposition rate was 0.2–0.3 nm/s. The amount of deposited Ag had an average thickness of 8 nm.

**2.3. Fabrication of Multilayer Thin-film Flakes Dispersion Gel.**

A 10% w/w hydroxyethyl cellulose (HEC) gel (SE400, Daicel FineChem Ltd.) was prepared in Milli-Q water. A 5 × 5 cm² ML1 and ML2 thin-film was cut into 5 mm squares and mixed put into a mixer with 9.0 cm² HEC gel and 40 stainless steel balls (SUS sphere number?, ϕ0.5 mm). The mixture was stirred using a centrifugal mixer (THINKY, AR-100). A schematic diagram of the stirring mechanism is shown in Fig. 1(c). The mixing container made a 45° angle with the vertical axis when placed inside the mixer. The centrifugal mixer simultaneously rotates and revolves in the opposite direction and creates vertical convection in the sample. During stirring, the rotational speed was set at 2000 rpm, and the revolution speed was 800 rpm. Several samples were prepared with increasing mixing time (1 to 10 min) to investigate the effect of mixing time on the size of the flakes to obtain flakes of desired thickness (10–100 μm).

**2.4. Morphological, Optical, and SERS Characterization.**

A scanning electron microscope (SEM; Hitachi High-Tech, SU3800) with a LaB6 detector at 10 kV acceleration voltage was used for the surface morphology study. A microscope (BM-3400TTR) at 10x magnification was used to test the prepared samples. Approximately 0.1 mm thick dispersion gel flakes were spread over a covering glass and capped with another covering glass. The area of thin-film flakes from microscopic images was calculated using image processing software ImageJ.[36] For SERS and reflectance measurements, a silicone cell was prepared, Fig. S1 (Supplementary Information). The cell had an area of 5×5 mm² and was 1.0 mm thick. A thin layer (approximately 0.5 mm) of the flake dispersion gel was filled in the cell. A 30 μL droplets of 1 mM aqueous 4,4-bipyridine (BPY) solution was deposited on the flake dispersion gel in the cell, and its SERS spectra were recorded as a function of time. SERS spectra were acquired using a Raman spectrometer (RAM200S; LambdaVision, Inc ). A 785 nm laser with a 50 × objective and 30 mW power on the sample was used for excitation.

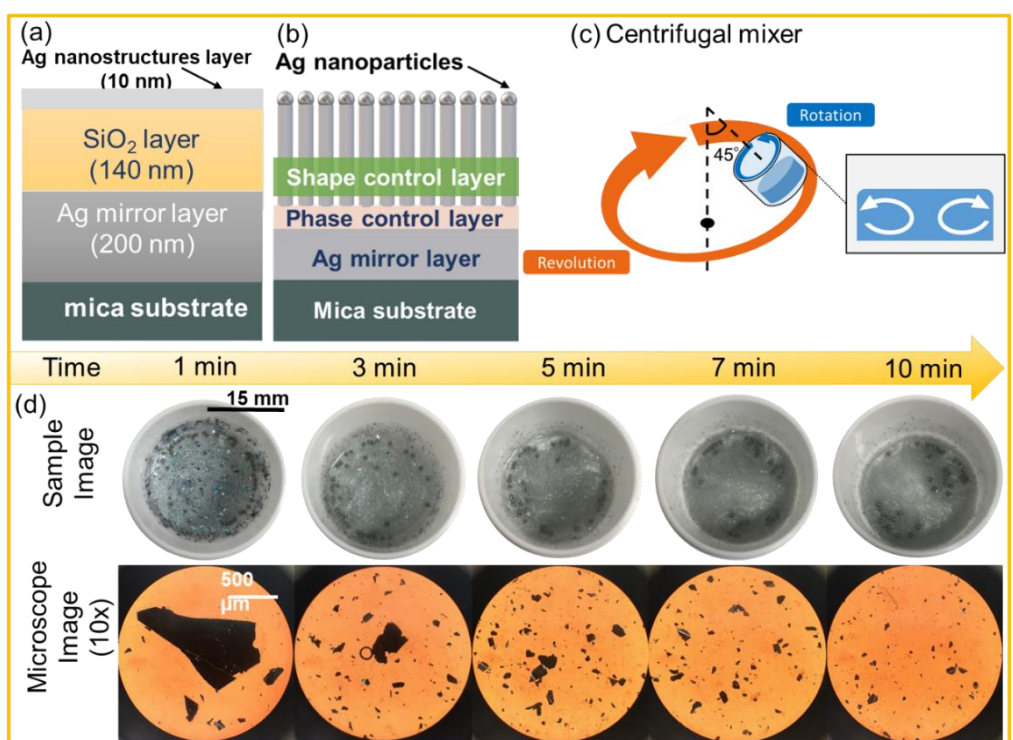

**Figure 1.** A schematic diagram of the cross section of (a) ML1; and (b) ML2 thin-films; (c) a schematic diagram of the centrifugal mixing. Arrows represent the state of the vertical and vertical flow generated in the sample; (d) actual photograph and microscopic image of the ML1 flake dispersion gel for different stirring time.

## 3. RESULTS AND DISCUSSION

### 3.1. The preliminary examination for the optimization of the size of thin-film flakes dispersed in the gel.

To produce a multilayer thin-film flakes gel composite, we first examined the target size of thin-film flakes from two perspectives: the surface morphology of thin-films and their effects on the human body. Figure 2 shows an SEM micrograph of the ML2 thin-film that we used for the SERS study. On the top layer, the Ag nanoparticles were aligned, representing anisotropic surface morphology, and the nanorod arrays were obtained with a large aspect ratio and surface nanogaps. The surface morphology of the thin-film determines its optical properties. The properties of the metal nanoparticles will change if they are dispersed, or when the particles themselves are disjointed, and the aspect ratio is changed. Therefore, thin-film flakes that maintain this surface structure should have an area of at least 1 µm². We used mica as a substrate for depositing multilayer thin-films as it can be processed thin enough compared to other substrate materials such as general glass and can maintain rigidity. In addition, the average particle diameter of fine mica particles used in commercial cosmetics is approximately 20–50 µm. Particles of the order of 10 µm are considered safe and have no detrimental effect on the human body, and the optical properties of thin-films can be reliably preserved. Therefore, in this study, we aimed to produce thin-film flakes with a particle size of 10-100 µm.

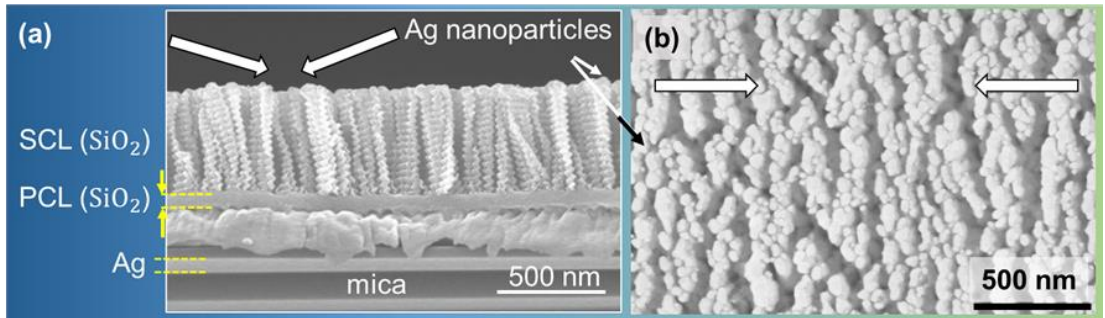

**Figure 2.** SEM micrograph of the ML2 thin-film. (a) cross-sectional; and (b) top view

Figure 1(d) shows the actual photograph and microscopic image of the ML1 flake dispersion gel. When the samples were stirred for more than 3 min, the thin-film flakes were so fine that the particles could not be observed with the naked eye. Moreover, there was no significant change in the color of the sample. As the stirring time increased, the number of thin-film flakes with large areas decreased. To calculate the particle size of thin-film flakes by image processing equivalent diameter ($l$) was defined as

$$l = 2\sqrt{\frac{S_i}{\pi}} \quad (1)$$

where $S_i$ is the area of a thin-film flake determined by image processing. Since a 10× objective lens with N.A. = 0.25 was used, the resolution in the visible light range was roughly 2.2 µm, and we assumed that particle sizes of at least 3.0 µm or more could be measured. Therefore, the minimum thin-film flakes with equivalent circle diameter 3 µm or more, that is, the area of approximately 7.1 µm² or more, could be detected in the particle size measurement.

The centrifugal force on the flakes during rotation will disperse the flakes of different particle sizes from the center of the container to the edge. Therefore, to calculate the area and particle size of thin-film flakes from microscopic images, we divided the samples into three regions: central, middle, and edge, with a radius of 0-5 mm, 5-10 mm, and 10-15 mm, respectively, Fig 3(b). We calculated the particle size using Eq.1. Figure 3(a) shows the relationship between the maximum particle size and the stirring time of thin-film flakes in each region. In 1-3 min, there was a large variation in the flakes size distribution, and flakes with larger size tended to appear at the edge, but after 4 min, such a trend was not observed, and the variation seemed to have converged to some extent. The particle size of the thin-film flakes decreased with increasing stirring time, and after stirring. For more than 4 min, all the particles were less than 500 μm without any large thin-film flakes. Hence, more than 4 min of stirring is needed to obtain more thin-film flakes of the target particle size. Next, we will take a closer look at how the amount of thin-film flakes in the target particle size changes after stirring for 4 min.

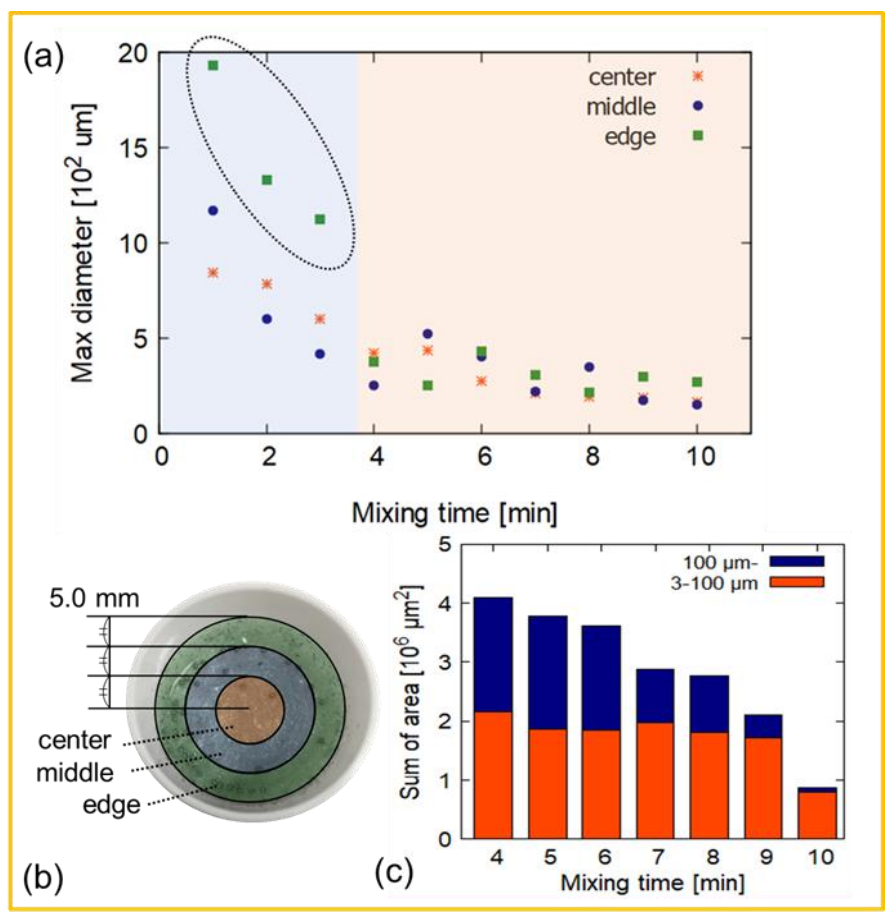

**Figure 3.** (a) Variation in maximum diameter of the dispersed flakes with stirring time; (b) schematic diagram of the division of regions for the particle size measurement; (c) dependence of the total area of thin-film flakes on stirring time. The total height of the bar represents the total area of each sample. The orange part represents the area of thin-film flakes with a particle size of 3-100 μm and the blue part with more than 100 μm.

The transition of thin-film flakes with particle sizes smaller than 100 μm in agitation for 4 min or more was investigated in more detail. Figure 3(c) shows the total area of the thin-film flakes with stirring time. The orange and blue bars are the total area of the thin-film flakes with particle sizes between 3-100 μm and 100 μm or more, respectively. The addition of the two represents the total area of the flakes in each sample. The total area of thin-film flakes with particle sizes between 3-100 μm was almost constant in 4-8 min, and the targeted thin-film flakes with a particle size of 10 μm order were obtained by stirring for 4-8 min. On the other hand, the total area of the thin-film flakes in the sample tends to decrease over time, and the total area of the thin-film flakes of the sample after 9 min was approximately half the sample after 4 min. However, this contradicts the result of the transmittance spectrum that the total area of the thin-film flakes contained in the sample was almost constant, Fig. S2 (Supplementary Information). This could be because the thin-film flakes were crushed into sizes smaller than that can be resolved using our setup. Since the percentage of particles with a size less than 100 μm increases after 6 min, we aimed to produce thin-film flakes with a particle size of 10-100 μm, we selected a stirring time of 4-6 min for the preparation of ML2 flake dispersion gel.

### 3.2. Ag NRA/SiO$_2$/SiO$_2$/Ag mirror multilayer thin-films.

Figure 2(a) shows the cross-sectional SEM image of the sample ML2 thin-film. The structure of the ML2 film consists of the AgNRAs layer, SCL and PCL using SiO$_2$, and an Ag mirror layer on mica from top to bottom, respectively. The white arrows in the figure indicate the deposition directions of Ag and SiO$_2$ flux during deposition. The film thickness of SCL was approximately 500 nm. Figure 4 shows the theoretical and experimental reflectance spectra of ML2 as a function of PCL thickness, $d_{PCL}$. The theoretical calculation of reflectance in provided in detail in the Supplementary Information. Striped patterns were observed, in which the reflectance changed periodically as a function of the photon energy at a specific value of $d_{PCL}$. The reflectance in the NIR region was much larger than that for the higher energies. At photon energy of ~1.4 eV, the reflectance becomes less than 1%. In addition, for photons with an energy of 1.58 eV, the reflectance was minimum for $d_{PCL}$ values between 60 and 140 nm.  Since we used a 785 nm laser for SERS, the absorption will be maximum for $d_{PCL}$= 60 to 140 nm and will show a strong SERS enhancement for that wavelength.

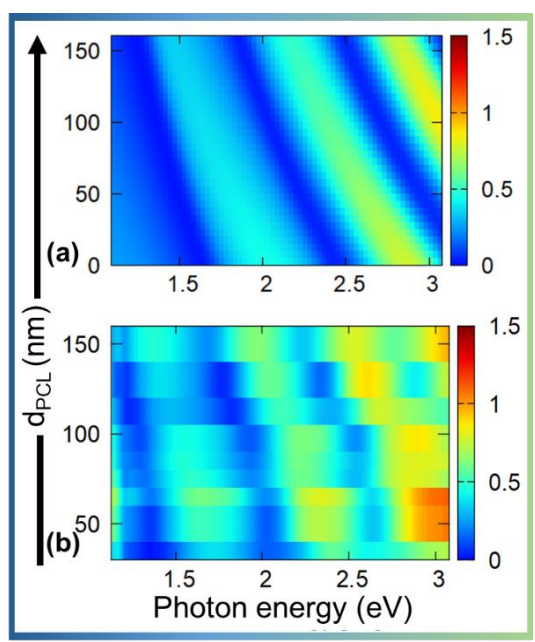

**Figure 4.** Calculated and measured reflectance of ML2 thin-film.

The SERS performance of the ML2 thin-film was studied along with the diffusion of the analyte through the gel for 60 nm ≤ $d_{PCL}$ ≤ 140 nm. To examine the characteristics of the multilayer film immersed in the gel, a cell, as shown in Fig. S1 (Supplementary Information) was prepared. ML2 thin-film was put inside the cell and coated with 0.5 mm of HEC gel. 30 μL of BPY solution was dropped on top of the gel, and SERS spectra were recorded from the bottom. The SERS spectra of BPY from the AgNRAs without the Ag mirror layer on mica and with PCL are shown in Fig. 5(a). The characteristic peaks of BPY at 1085 cm$^{-1}$, 1240 cm$^{-1}$, and 1509 cm$^{-1}$ can be assigned to the A1 ring mode. The B2 ring modes appeared at 1300 cm$^{-1}$, and 1624 cm$^{-1}$.[37] BPY molecules diffused through the gel rapidly and reached the Ag nanoparticles in less than 60 s, giving Raman peaks Fig S3 (Supplementary Information). Meanwhile, the process was driven by diffusion; the Raman signal increased continuously with time and was almost constant after 20 min. SERS data was taken after 60 min for our study when the Raman peak was well stabilized. In all the ML2 thin-film samples multilayer film, a strong SERS peak was obtained than the AgNRAs. This is because of the coupling of the light to the local plasmons in the Ag nanorods was successfully modified by interference with the Ag mirror surface in ML2 thin-films. SERS intensity was maximum for $d_{PCL}$= 105 nm. When the thickness of the PCL was 105 nm was maximum. The peak intensity was roughly seven times that of the Ag nanorod array on the mica substrate. The relationship between the 1624 cm$^{-1}$ peak intensity and PCL film thickness measured after 60 minutes from is shown in Fig. 6(b). It was confirmed that the difference in PCL film thickness could modify the absorbance and enhance the SERS performance of the AgNRs. Additionally, the enhancement was maximum for ML2 thin film with $d_{PCL}$=105 nm.

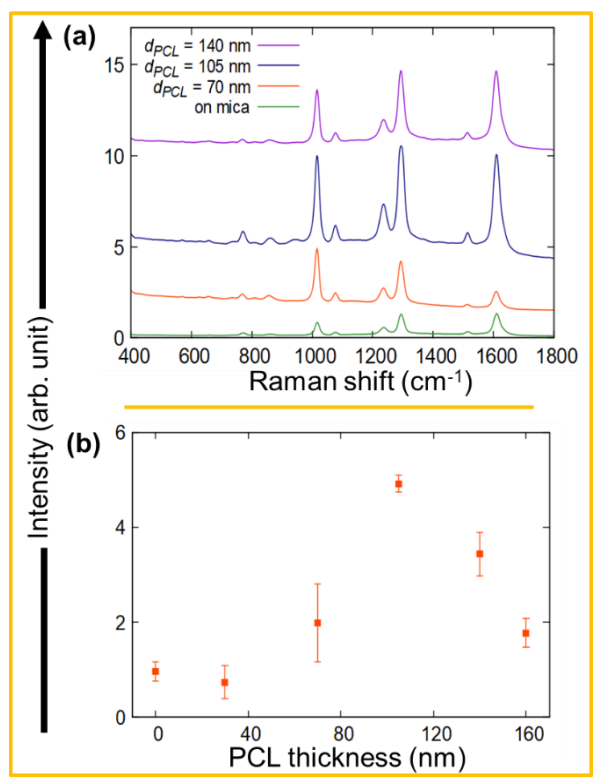

**Figure 5.** (a) SERS spectra of BPY on ML2 thin-film with $d_{PCL}$=70 nm, 105 nm, 140 nm, and AgNRAs on a mica substrate in gel of thickness 0.5 mm. The data was recorded after 60 min of dropping BPY solution; (b) SERS peak intensity (1624 cm$^{-1}$) vs PCL film thickness.

### 3.3. Ag NRA/SiO$_2$/SiO$_2$/Ag Mirror Multilayer Thin-film Dispersion Gel.

Finally, we prepared a thin-film flake dispersion gel with ML2 thin-film using the method discussed in Section 3.1. Figure 7(a) shows an ordinary photograph of the flake dispersion gel before and after stirring for 5 min. The ML2 dispersion gel with $d_{PCL}$ = 70 nm and 105 nm was uniform and transparent, whereas the samples using AgNRAs had a significant change in appearance compared to before mixing and were dark in color. The change in color could be attributed to the damage of Ag nanorod array on the surface of the thin-film flake and change in the aspect ratio. A small change in the aspect ratio can lead to a drastic change in the transmitted color of the nanoparticles.[38]
To examine the sample in more detail, ML2 flakes dispersion gel of thickness 1 mm was observed with a microscope using a 10× objective lens. The microscopic image of the sample after stirring for 5 minutes is shown in Fig. 6. In the sample using ML2 thin-film flakes with the same particle size, as discussed in section 3.1 were obtained. On the other hand, in the AgNRAs thin-film, the number of recognizable thin-film flakes decreased. This may be because the AgNRAs were peeled off from the mica substrate and cracking of AgNRAs to a smaller size than the resolution of the microscope. The AgNRAs thin-film does not have a PCL or silver mirror surface. Therefore, we assume that the PCL and Ag mirror layer maintains the original thin-film structure even after the film is crushed.

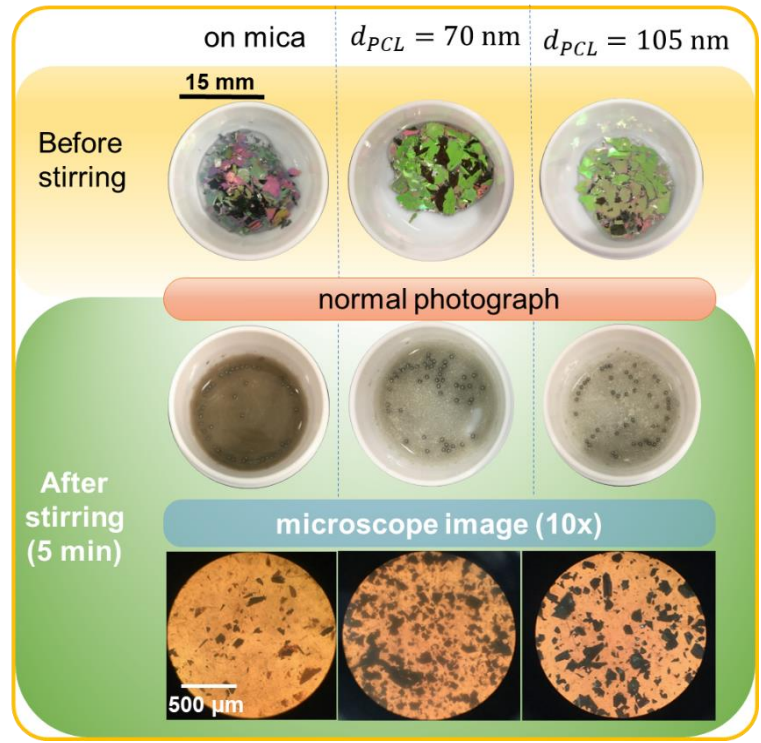

**Figure 6.** Normal photograph and microscopic image of the ML2 thin-film dispersion gel, with different $d_{PCL}$ thicknesses, before and after stirring.

Figure 7 shows the Raman spectrum of each sample obtained by SERS measurement. SERS measurements were taken after 5 min of stirring, and measurements were performed after 60 min of dropping BPY after the SERS signal was stabilized. The characteristic peaks of BPY were observed in each sample. The SERS measurements were performed at multiple points on each sample by changing the stirring time. Figure 7(b) shows a plot of the three highest peak intensities for a specific stirring time. When the stirring time was about 4-8 min, ML2 thin-film flakes dispersion sample with $d_{PCL}$ = 105 nm showed a larger peak than the other samples. In particular, the ML2 samples with $d_{PCL}$ = 105 nm after 5 min and 7 min showed roughly 7 times the enhancement compared with the samples using AgNRAs thin-film only. The peak intensity of the AgNRAs thin-film was small and remained almost constant regardless of the stirring time. Therefore, a thin-film flake dispersion gel with a multilayer film structure of the Ag nanorod array/dielectric/silver mirror surface showed the best SERS enhancement.

The SERS measurements were taken at various positions by moving the sample cell in the in-plane direction. There were many spots where the SERS peak could not be obtained. Image processing was performed using the microscope image and ImageJ to determine the number density and average area of the dispersed flakes. The number density was $7.5 \times 10^{-8}$ μm$^{-3}$, and the average area per thin-film flake was $5.6 \times 10^3$ μm$^2$. From these, the average distance between flakes was estimated to be ~ 640 μm. The laser spot diameter used for Raman spectroscopy in this study, on the other hand, was 1.6 μm as obtained from the wavelength and numerical aperture of the objective lens for an ideal focusing. In other words, the distance between the thin-film flakes was very large with respect to the laser spot diameter. Ideally, the spot diameter should be less than the distance between the thin-film flakes in order to reduce the measurement variability. The depth of focus of the objective lens used in the Raman spectroscope used in this study was ± 3.0 μm, and that this depth may also affect the thin-film flake detection range. This problem can be solved by increasing the flake density by approximately 10 times. Additionally, if it is possible to select a laser light source and an optical system with a depth of focus that can include multiple thin-film flakes, SERS performance of the ML2 thin-film flake dispersion gel can be further improved.

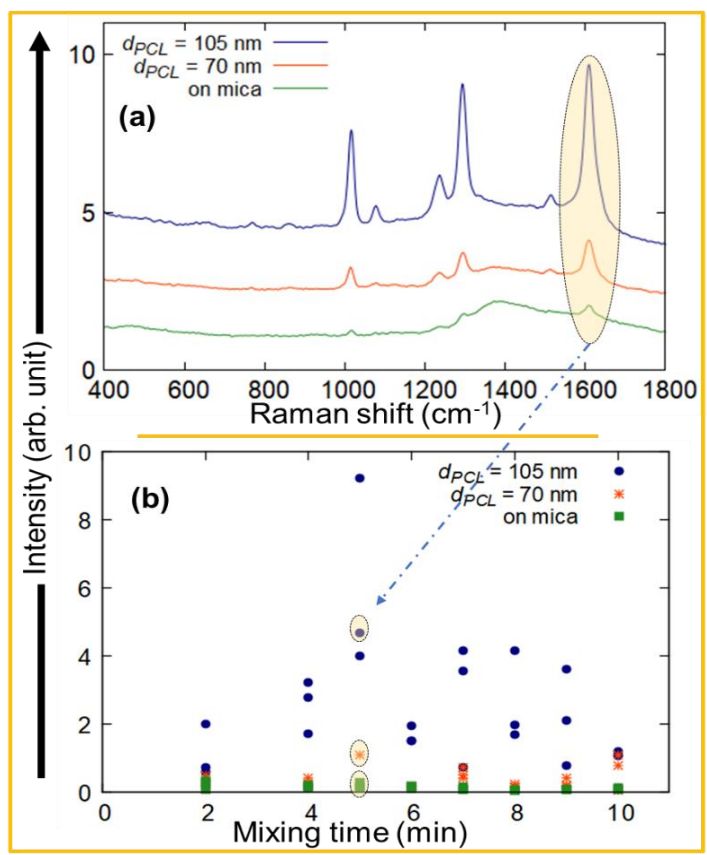

**Figure 7.** (a) SERS spectra of ML2 thin-film ($d_{PCL}$ = 70 nm, and 105 nm) and AgNRAs dispersion gel after stirring for 5 min; (b) SERS peak intensity (1624 cm$^{-1}$) vs. stirring time ML2 thin-film ($d_{PCL}$ = 70 nm, and 105 nm) and AgNRAs dispersion gel. The data was recorded after 60 min of dropping BPY solution.

## 4. CONCLUSIONS

In summary, we fabricated an Ag NRA/SiO$_2$/SiO$_2$/Ag mirror multilayer thin-film dispersion gel SERS substrate by centrifugal mixing of the multilayer thin-film in HEC gel. The thin-film flakes with the particle size 10-100 μm can be obtained by stirring for 4-6 min that retains the properties of the original thin-film and has high performance. It was also found that stirring for more than 7 min improved the ratio of thin-film flakes of the target particle size, but the characteristics of the original thin-film may not be retained, and the performance may deteriorate. Multilayer thin film with $d_{PCL}$ = 105 nm showed the best SERS enhancement. The SERS signal was enhanced approximately seven times than the Ag nanorod array substrates.

**SUPPLEMENTARY MATERIALS:** The following are available online: Figure S1: Schematic diagram of a sample cell used for reflectance and SERS measurement, Figure S2: Transmittance vs. mixing time for ML1 thin-film flakes dispersion gel, Figure S3: SERS spectra in the gel for ML2 thin film with $d_{PCL}$ = 105 nm vs. time, theoretical calculation of reflectance of ML2 thin-film.

**FUNDING:** This work was supported by JST COI Grant Number JPMJCE.1307.

**ACKNOWLEDGMENTS:** The authors also thank Dr. Kosuke Ishikawa of Kyoto University for assisting us with the SEM observations.

**CONFLICT OF INTEREST:** The authors declare no conflict of interest.

# REFERENCES

1 A. B. and P. N. Vladimir Bochenkov, Jeremy Baumberg, Mikhail Noginov, Felix Benz, Hasan Aldewachi, Silvan Schmid, Viktor Podolskiy, Javier Aizpurua, Kaiqiang Lin, Thomas Ebbesen, Alexei A Kornyshev, James Hutchison, Katarzyna Matczyszyn, Samir Kumar, Bart de Nijs, Franci, *Faraday Discuss.*, 2015, **178**, 435.

2 K. M. Kosuda, J. M. Bingham, K. L. Wustholz, R. P. Van Duyne and R. J. Groarke, *Nanostructures and Surface-Enhanced Raman Spectroscopy*, Elsevier Ltd., 2016.

3 S. Kruszewski, in *Proc. SPIE 3320, Tenth Polish-Czech-Slovak Optical Conference: Wave and Quantum Aspects of Contemporary Optics*, International Society for Optics and Photonics, 1998, vol. 3320, p. 281.

4 C. Y. Deng, G. L. Zhang, B. Zou, H. L. Shi, Y. J. Liang, Y. C. Li, J. X. Fu and W. Z. Wang, *Adv. Mater. Res.*, 2013, **760–762**, 801–805.

5 M. Moskovits, *Rev. Mod. Phys.*, 1985, **57**, 783–826.

6 S. Kumar, P. Kumar, A. Das and C. S. Pathak, in *Raman Scattering*, ed. K. P. Kumar Samir, IntechOpen, 2020.

7 J. Hughes, E. L. Izake, W. B. Lott, G. A. Ayoko and M. Sillence, *Talanta*, 2014, **130**, 20–25.

8 K. Sivashanmugan, K. Squire, A. Tan, Y. Zhao, J. A. Kraai, G. L. Rorrer and A. X. Wang, *ACS Sensors*, 2019, **4**, 1109–1117.

9 S. Kumar, D. K. Lodhi, P. Goel, N. Neeti, P. Mishra and J. P. Singh, *Chem. Commun.*, 2015, **51**, 12411–12414.

10 M. M. Harper, K. S. McKeating and K. Faulds, *Phys. Chem. Chem. Phys.*, 2013, **15**, 5312–28.

11 R. Petry, M. Schmitt and J. Popp, *Chemphyschem*, 2003, **4**, 14–30.

12 H. Pu, W. Xiao and D. W. Sun, *Trends Food Sci. Technol.*, 2017, **70**, 114–126.

13 S. Schlücker, *Angew. Chemie - Int. Ed.*, 2014, **53**, 4756–4795.

14 J. Yang, D. Shen, L. Zhou, W. Li, J. Fan, A. M. El-Toni, W. X. Zhang, F. Zhang and D. Zhao, *Adv. Healthc. Mater.*, 2014, **3**, 1620–1628.

15 Y. Shen, X. Cheng, G. Li, Q. Zhu, Z. Chi, J. Wang and C. Jin, *Nanoscale Horizons*, 2016, **1**, 290–297.

16 M. Fan, G. F. S. Andrade and A. G. Brolo, *Anal. Chim. Acta*, 2011, **693**, 7–25.

17 K. Xu, C. Zhang, T. H. Lu, P. Wang, R. Zhou, R. Ji and M. Hong, *Guangdian Gongcheng/Opto-Electronic Eng.*, 2017, **44**, 185–191.

18 N. A. Cinel, S. Cakmakyapan, S. Butun, G. Ertas and E. Ozbay, *Photonics Nanostructures - Fundam. Appl.*, 2015, **15**, 109–115.

19 M. Suzuki, W. Maekita, Y. Wada, K. Nakajima, K. Kimura, T. Fukuoka and Y. Mori, *Appl. Phys. Lett.*, 2006, **88**, 203121.

20 S. Kumar, P. Goel and J. P. J. P. Singh, *Sensors Actuators B Chem.*, 2017, **241**, 577–583.

21 M. Suzuki, W. Maekita, Y. Wada, K. Nakajima, K. Kimura, T. Fukuoka and Y. Mori, 2005, vol. 900, pp. 217–222.

22 M. Suzuki, Y. Imai, H. Tokunaga, K. Nakajima, K. Kimura, T. Fukuoka and Y. Mori, in *Proc. SPIE 7041, Nanostructured Thin Films*, 2008, vol. 7041.

23 S. Kumar, D. K. Lodhi and J. P. Singh, *RSC Adv.*, 2016, **6**, 45120–45126.

24 M. Suzuki, W. Maekita, Y. Wada, K. Nagai, K. Nakajima, K. Kimura, T. Fukuoka and Y. Mori, *Nanotechnology*, 2008, **19**, 265304.

25 M. J. Banholzer, J. E. Millstone, L. Qin and C. A. Mirkin, *Chem. Soc. Rev.*, 2008, **37**, 885–897.

26    P. A. Mosier-Boss, *Nanomaterials*, 2017, **7**, 142.

27    B. Doherty, F. Presciutti, A. Sgamellotti, B. G. Brunetti and C. Miliani, *J. Raman Spectrosc.*, 2014, **45**, 1153–1159.

28    S. Yao, C. Zhou and D. Chen, *Chem. Commun.*, 2013, **49**, 6409–6411.

29    K. Shin, K. Ryu, H. Lee, K. Kim, H. Chung and D. Sohn, *Analyst*, 2013, **138**, 932–938.

30    S. Kumar, K. Namura and M. Suzuki, *Proc. SPIE*, 2019, **10894**, 1089414.

31    J. G. Gibbs, A. G. Mark, T.-C. Lee, S. Eslami, D. Schamel and P. Fischer, *Nanoscale*, 2014, **6**, 9457–66.

32    S. Lucht, T. Murphy, H. Schmidt and H.-D. Kronfeldt, *J. Raman Spectrosc.*, 2000, **31**, 1017–1022.

33    S. Kumar, P. Goel, D. P. Singh and J. P. Singh, *Appl. Phys. Lett.*, 2014, **104**, 023107.

34    K. Namura, S. Imafuku, S. Kumar, K. Nakajima, M. Sakakura and M. Suzuki, *Sci. Rep.*, 2019, **9**, 4770.

35    S. Kumar, Y. Doi, K. Namura and M. Suzuki, *ACS Appl. Bio Mater.*, 2020, **3**, 3226–3235.

36    W. . Rasband, .

37    T. C. Strekas and P. S. Diamandopoulos, *J. Phys. Chem.*, 1990, **94**, 1986–1991.

38    P. K. Jain, K. S. Lee, I. H. El-Sayed and M. A. El-Sayed, *J. Phys. Chem. B*, 2006, **110**, 7238–7248.